\documentclass[twocolumn,pre,showpacs,superscriptaddress]{revtex4}

\usepackage{graphicx}
\usepackage{amsmath}
\usepackage{color}

\begin{document}

\title{Area coverage of radial L\'{e}vy flights with periodic
boundary conditions}

\author{Mahsa Vahabi}
\affiliation{Physics Department, Technical University of Munich,
D-85747 Garching, Germany}
\affiliation{Department of Physics,
Shahid Beheshti University, G. C., Evin, Tehran 19839, Iran}
\author{Johannes H. P. Schulz}
\affiliation{Physics Department, Technical University of Munich,
D-85747 Garching, Germany}
\author{Babak Shokri}
\affiliation{Department of Physics, Shahid Beheshti University, G.
C., Evin, Tehran 19839, Iran} \affiliation{Laser-Plasma Research
Institute, Shahid Beheshti University, G. C., Evin, Tehran 19839,
Iran}
\author{Ralf Metzler}
\affiliation{Physics Department, University of Potsdam, D-14476 Potsdam-Golm,
Germany}
\affiliation{Physics Department, Tampere University of Technology, FI-33101
Tampere, Finland}

\begin{abstract}
We consider the time evolution of two-dimensional L{\'e}vy flights in a
finite area with periodic boundary conditions. From simulations we show
that the fractal path dimension $d_f$ and thus the degree of area
coverage grows in time until it reaches the saturation value $d_f=2$ at
sufficiently long times.
We also investigate the time evolution of the probability density function
and associated moments in these boundary
conditions. Finally we consider the mean first passage
time as function
of the stable index. Our findings are of interest to assess the ergodic
behavior of L{\'e}vy flights, to estimate their
efficiency as stochastic search
mechanisms and to discriminate them from other types of search processes.
\end{abstract}


\maketitle

\section{Introduction}

L{\'e}vy flights are random walk processes, in which the lengths of individual
jumps are distributed according to a probability density $\lambda(x)$ of the
asymptotic power-law form
\cite{mandelbrot,hughes,bouchaud,report}
\begin{equation}
\label{jld}
\lambda(x)\simeq\frac{\sigma^{\alpha}}{|x|^{1+\alpha}},\,\,\,0<\alpha<2,
\end{equation}
where $\sigma$ is a scaling factor of physical dimension length,
$[\sigma]= \mathrm{cm}$. The resulting motion is spatially
scale-free due to the divergence of the jump length variance, $\int x^2
\lambda(x)dx$.
L{\'e}vy flights were popularized by Beno\^{\i}t Mandelbrot, who
named them after his teacher, French mathematician Paul L{\'e}vy
\cite{levy}. They have been mainly applied in the modeling of
search processes, following the original idea by Klafter and
Shlesinger \cite{klashle}: while regular random walks in one or two
dimensions have a high probability to return to already visited
sites, L{\'e}vy flights combine thorough local search with
occasional long excursions, leading them to areas which likely have
not been visited before. This strategy reduces unnecessary
oversampling and thus represents an advantage to the searcher.
Indeed, trajectories consistent with the long-tailed jump length
pattern (\ref{jld}) have been reported for various animal species
\cite{ghandi,mateos,sims,bartumeus}. Even for molecular search, L{\'e}vy flights
may emerge from the topology of the search space \cite{michael}. Due to a mix
of modern means of transportation also human motion behavior is
characterized by travel lengths with power-law distribution
\cite{brockmann}. Long-tailed distributions of relocation lengths
also change significantly the distribution pattern of diseases, as
regular diffusion fronts are broken by, for instance, long-distance
air travel, thus carrying the disease to completely decoupled places
\cite{brockmann1}.

For land-based animals typical search processes are essentially
two-dimensional. In many cases we may also neglect the vertical
dimension for many birds or fishes, when the lateral extension is
considerably larger than the maximal height/depth difference of the
trajectory. In such a two-dimensional search space, given its
fractal dimension $d_f=\alpha$ a single L{\'e}vy flight trajectory
cannot fully cover the search space in absence of boundaries. Thus,
some small area element in the search space may be hit by one single
trajectory while it
may be missed by another. In many cases, however, the search space
is bounded. For instance, animals only search for food in their own
territory, or the business travel patterns of an individual are
confined to a certain country or continent. Moreover it is often
relevant to have information about the area coverage of a L{\'e}vy
flight process at finite times. For instance, a predator often does
not cover its entire territory on a single day. It is therefore of
interest to explore the time evolution of the area coverage of
L{\'e}vy flights: how long does it take for the animal to
efficiently explore its entire patch, or a disease to reach every
little town in a country? How large is the area coverage at a given
time? These questions of area coverage are directly connected with
the ergodic properties of L{\'e}vy flights.

Incomplete area coverage of L{\'e}vy flights is also of interest in view
of their ergodic properties. Ergodicity and its violation have recently
received considerable interest, following the possibility to record single
trajectories of molecules or small tracers in biological matter \cite{pt}.
Thus, for anomalous diffusion processes with underlying scale-free motion
a weak ergodicity breaking is observed \cite{weigel,lene}, such that the
mean squared displacement obtained from the time average over a single
trajectory is a random variable and significantly different from the
corresponding ensemble average, as shown for the subdiffusive continuous
time random walk process \cite{he,pccp,pnas,lubelski,igor_epl}.
This behavior contrasts the ergodic behavior of other anomalous diffusion
processes such as fractional Brownian motion \cite{deng,jae_prl}. While
weak ergodicity violation occurs for processes with diverging characteristic
time scales, ergodicity is also violated for L{\'e}vy flights in infinite
media, due to the incomplete area coverage even for infinitely long
trajectories \cite{lutz}. Confined L{\'e}vy flights, in contrast, are ergodic
\cite{marcin}.

In this paper, we consider radial L{\'e}vy flights in two-dimensional
square areas with periodic boundary conditions. We investigate the
time evolution of the effective fractal dimension of the flights for
varying stable index $\alpha$. As a complementary measure for the
area coverage we analyze the time evolution of the mean squared
displacement up to its saturation. Moreover we analyze the time
evolution of the probability density function and its
moments. We also pursue the question on the typical time for the
L{\'e}vy flight to first reach or pass the boundary of the square
interval.

The paper is organized as follows. In section II, we briefly review the
mathematical foundation of L{\'e}vy flights.
In section III, we consider radial two-dimensional L{\'e}vy flights in square
boxes and numerically study their area coverage in terms of their fractal
dimension, their probability density function, evolution of moments, and
mean first passage times. Section IV summarizes the results.

\section{L\'{e}vy flights}

\begin{figure}
\includegraphics[width=8cm]{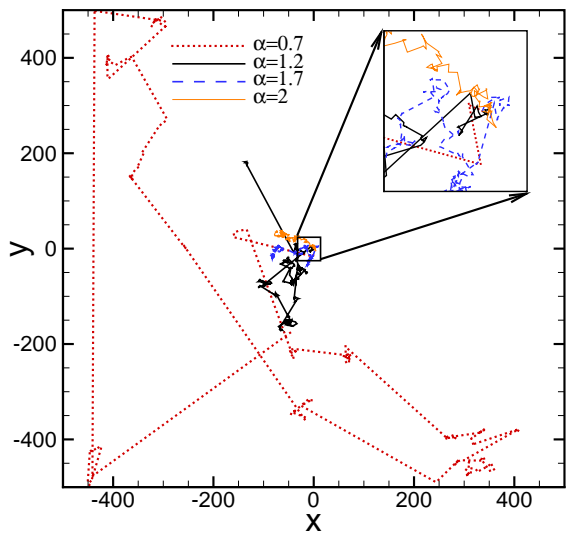}
\caption{(Color online)
Trajectories of radial two-dimensional L{\'e}vy flights for different
values of the stable exponent $\alpha$, showing 400 steps each in a $1000
\times1000$ box under periodic boundary conditions. Starting point was the
center of the square area. Note the significantly longer excursions for
smaller values of $\alpha$. We chose $\sigma=\sqrt{2}$.}
\label{fig.trajectory}
\end{figure}

For jump length distributions of the type~(\ref{jld}), the variance
$\langle x^2\rangle=\int_{-\infty}^{\infty}x^2\lambda(x)dx$ diverges,
while fractional moments $\langle|x|^{\delta}\rangle$ of order $0<\delta<
\alpha$ exist \cite{report}. This property directly carries over to the
random motion itself and is responsible for several peculiar phenomena.
Due to this lack of a finite variance L{\'e}vy flights have a fractal
graph dimension $d_f=\alpha$ \cite{hughes}. This property causes the
sample path of a L{\'e}vy flight to be characterized by occasional long jumps,
connecting clusters of more localized jumps. Unvisited holes exist on all
scales within the sample path. This scale-free behavior gives rise to the
fact that clusters host smaller clusters, as shown in
Fig.~\ref{fig.trajectory}. Moreover, the discontinuous jumps on arbitrarily
large scales induce a principal discrepancy between first passage and first
arrival events. For instance, while
their first passage behavior on a semi-infinite domain obeys the Sparre
Anderson universality \cite{chechkin,koren}, their first arrival behavior
is significantly reduced with decreasing $\alpha$ \cite{chechkin}, as L{\'e}vy
flights strongly overshoot a point target \cite{koren}. On finite domains the
first passage behavior is also modified \cite{Dybiec,Condamin}. Another
interesting effect
is the occurrence of multimodal distributions for L{\'e}vy flights in the
presence of steeper than harmonic potentials \cite{chechkin1,jespersen}.

Analytically, the statistics of the L{\'e}vy flight are dictated by the
detailed distribution of individual jump lengths. However, on large
scales---when large search spaces are explored for sufficiently long periods
of time---we obtain a universal description. Displacements on large time scales
are characterized by L{\'e}vy $\alpha$-stable laws, which in the
symmetric case are defined in terms of their characteristic function $\langle
\exp(ikx)\rangle$
\cite{levy,gnedenko,hughes,bouchaud,report},
\begin{equation}
\label{char_func}
\lambda(k)\equiv\int_{-\infty}^{\infty}\lambda(x)e^{ikx}dx=\exp\left(-
\sigma^{\alpha}|k|^{\alpha}\right).
\end{equation}
Again, $\sigma$ is a typical
length scale. From the details of the microscopic jump length distribution,
only the tail property~(\ref{jld}) is directly passed on to the asymptotic
stable law~(\ref{char_func}) of macroscopic displacements, including in
particular the exact tail exponent $\alpha$.
L{\'e}vy stable laws emerge as the limit of the sum of independent, identically
distributed random variables with diverging variance, by virtue of the
generalized central limit theorem \cite{levy,gnedenko,hughes,bouchaud}.
For $0<\alpha<2$ this leads to the asymptotic power-law (\ref{jld}), while
in the limit $\alpha=2$, we recover a Gaussian distribution for $\lambda(x)$
with finite moments of all orders.

Here, we consider two-dimensional L{\'e}vy flights in square boxes
of edge length $2a$, with periodic boundary conditions. This means
that each time the particle crosses one edge of the box, it will
enter from the opposite edge. All L{\'e}vy flights start from the
origin in the center of the square box. For each step in the two
dimensional random walk we draw a (signed, symmetric) flight
distance $r$, with distribution $\lambda(r)$ as in
(\ref{char_func}), and independently a flight direction $\theta$,
uniformly distributed within $[0;\pi]$. The projection of
displacements onto the $x$-axis is thus $\Delta x = r\cos(\theta)$. We call
such processes \emph{radial\/} two-dimensional L{\'e}vy flights. Such radial
L{\'e}vy flights have been used to analyze search processes (see, e.g.,
Ref.~\cite{ghandi}). We note that they are slightly different from
two-dimensional L{\'e}vy flights defined in terms of the two-dimensional
characteristic function $\exp(-[\sigma^*]^{\alpha}|\mathbf{k}|^{\alpha})$
\cite{chechkin_gonchar}.

In what follows we calculate parameters such as the fractal
dimension, the mean first passage time, the mean squared
displacement, the probability density function and its other moments
and follow their temporal evolution.

\section{Results and discussion}

\subsection{Fractal dimension}

The fractal dimension $d_f$ indicates how completely a fractal graph fills
the available embedding space. For a L{\'e}vy flight evolving in our finite
area the value of the fractal dimension will initially be smaller than two.
Starting from the value zero for the first (few) point(s) of the trajectory,
if the area is sufficiently large the value of $d_f$ is expected to be
the value of the stable index, $d_f=\alpha$, as long as the boundary
effects do not come into play. In our simulations we do not see a saturation
close to $d_f=\alpha$, as we are interested in the finite size effects and
therefore choose quite small domains. Due to this finiteness of the domain the
L{\'e}vy flight will eventually cover the entire area and thus the fractal
dimension is expected to saturate at the value $d_f=2$. This corresponds
to the observation of ergodicity for bounded L{\'e}vy flights \cite{marcin}.

\begin{table}
\begin{tabular}{|c|c|c|}
\hline
Input $\alpha$ & Number of steps & Output $\alpha$ \\\hline
$0.7$    & $10,000$         & $0.68 \pm 0.06$ \\\hline
$1.2$    & $50,000$         & $1.16 \pm 0.04$ \\\hline
$1.7$    & $10,000,000$      & $1.65 \pm 0.03$ \\\hline
\end{tabular}
\caption{\label{Tb1} Value of the stability index $\alpha$ of free
L\'{e}vy flights. We list the input $\alpha$ used to generate the
trajectory and the output value for $\alpha$ as determined from the
box counting method. The number of steps is an indication for the
number of steps of the simulated L{\'e}vy flight necessary to
determine $\alpha$ to reasonable accuracy.}
\end{table}
To determine the time evolution of the fractal dimension of the L{\'e}vy flight
under consideration we calculate the box counting dimension of the trajectory
\cite{falconer}. In this method, the number of boxes $N$ containing a point of
the fractal set (here, the points of visitation of our trajectory)
is counted for a given box size $\epsilon^2$. For an underlying
fractal geometry and sufficiently small $\epsilon$ a power law relation of
the form $N(\epsilon)\simeq\epsilon^{-d_f}$ is expected \cite{falconer}.
To validate our method we first determined the fractal dimension of a number
of unbounded L{\'e}vy flights, for which the value $d_f=\alpha$ is expected.
The results are summarized in Tab.~\ref{Tb1}. We find that the fractal
dimension is reproduced quite reliably. However, the number of necessary
random walk steps to obtain sufficiently good results increases dramatically
with growing $\alpha$.

\begin{figure}
\includegraphics[width=8cm]{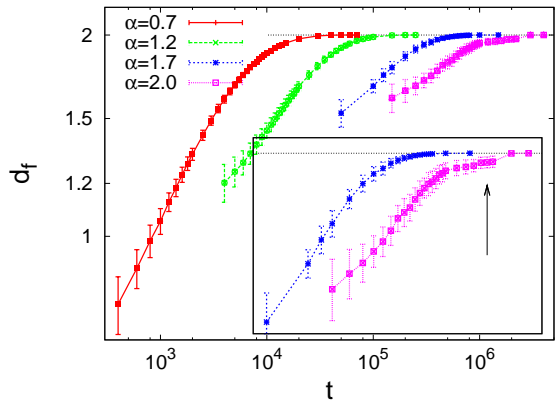}
\caption{Fractal dimension ($d_f$) of L{\'e}vy flights on the square domain
with periodic boundary conditions as function of time for different stable
indices $\alpha$, with $\sigma=\sqrt{2}$ and $a=500$. Here, $200$ sample
trajectories
were used for the ensemble average to produce $d_f$ at a given time $t$. The
inset magnifies the plateau region for $\alpha=2$ compared to the case
$\alpha=1.7$.}
\label{fig.fracdim}
\end{figure}

In Fig.~\ref{fig.fracdim}, we plotted the time-dependent fractal
dimension calculated via the two-dimensional box-counting method for
confined L{\'e}vy flights with different values of $\alpha$
($\alpha=0.7$, $1.2$, $1.7$, and $2$) for fixed box size
$1,000\times1,000$. As expected for all $\alpha$ values the fractal
dimension reaches the Euclidean embedding dimension, $d_f\to2$. We
see that for growing $\alpha$ indeed the slope of the curve $d_f(t)$
is smaller, such that the stationary limit is reached slower.
Interestingly, for the Gaussian limit $\alpha\to2$ an intermediate
plateau emerges, as indicated by the arrow in the inset of
Fig.~\ref{fig.fracdim}. By analysis of a large number of L{\'e}vy
flights for this $\alpha$ value we made sure that this effect is not
an artifact. The significance of this observation is also underlined
by the relatively small statistical sample error bars shown in
Fig.~\ref{fig.fracdim}. Thus, close to $\alpha=2$ care should be
generally taken when determining the fractal graph dimension of a
L{\'e}vy flight process.

The relaxation time to the ergodic behavior of full area coverage of a L{\'e}vy
flight may reach considerable values. For a searching animal it is therefore
imperative to have a reasonable field of vision, i.e., to be able to scan an
area around each point of the trajectory. Even more efficient is constant
scanning as discussed in Ref.~\cite{ghandi}, such that the scanned on-the-fly
area represents a sausage. Without such provisions the scanning of the area
by the L{\'e}vy flight remains sparse \emph{a forteriori}, unless extremely
long search times are invested.

\subsection{Probability density function}

\begin{figure}[]
\includegraphics[width=8cm]{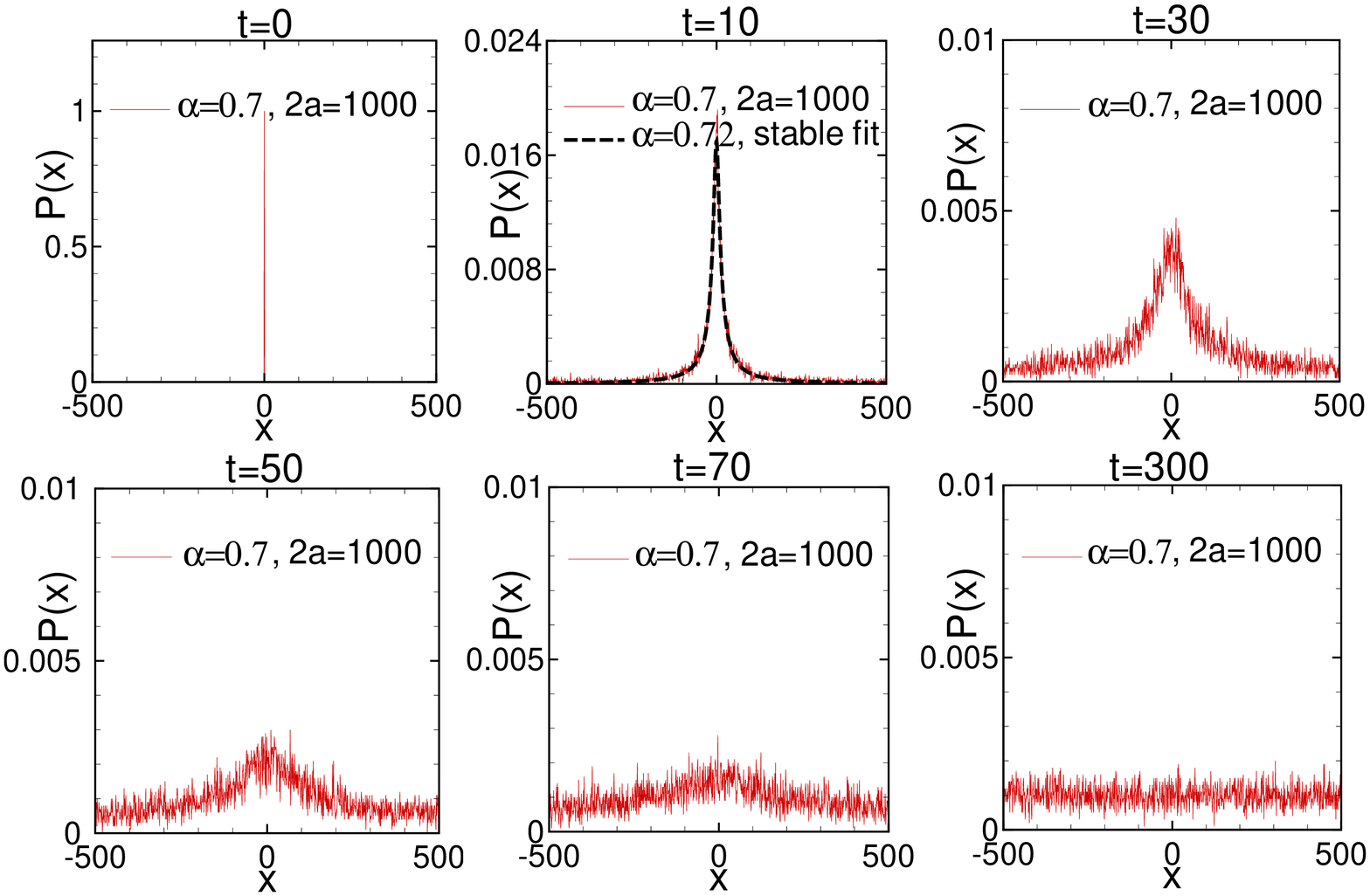}
\includegraphics[width=8cm]{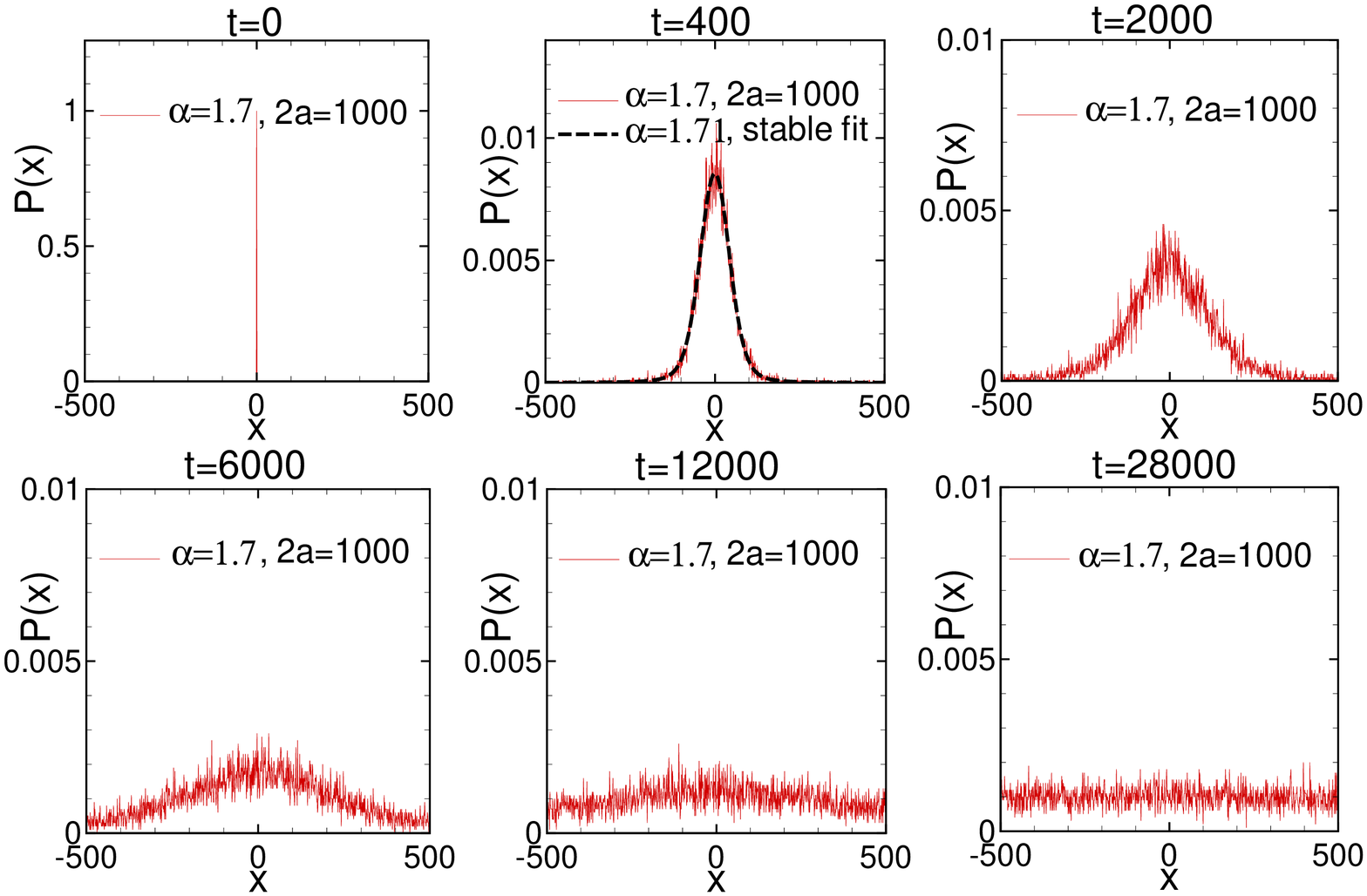}
\includegraphics[width=8cm]{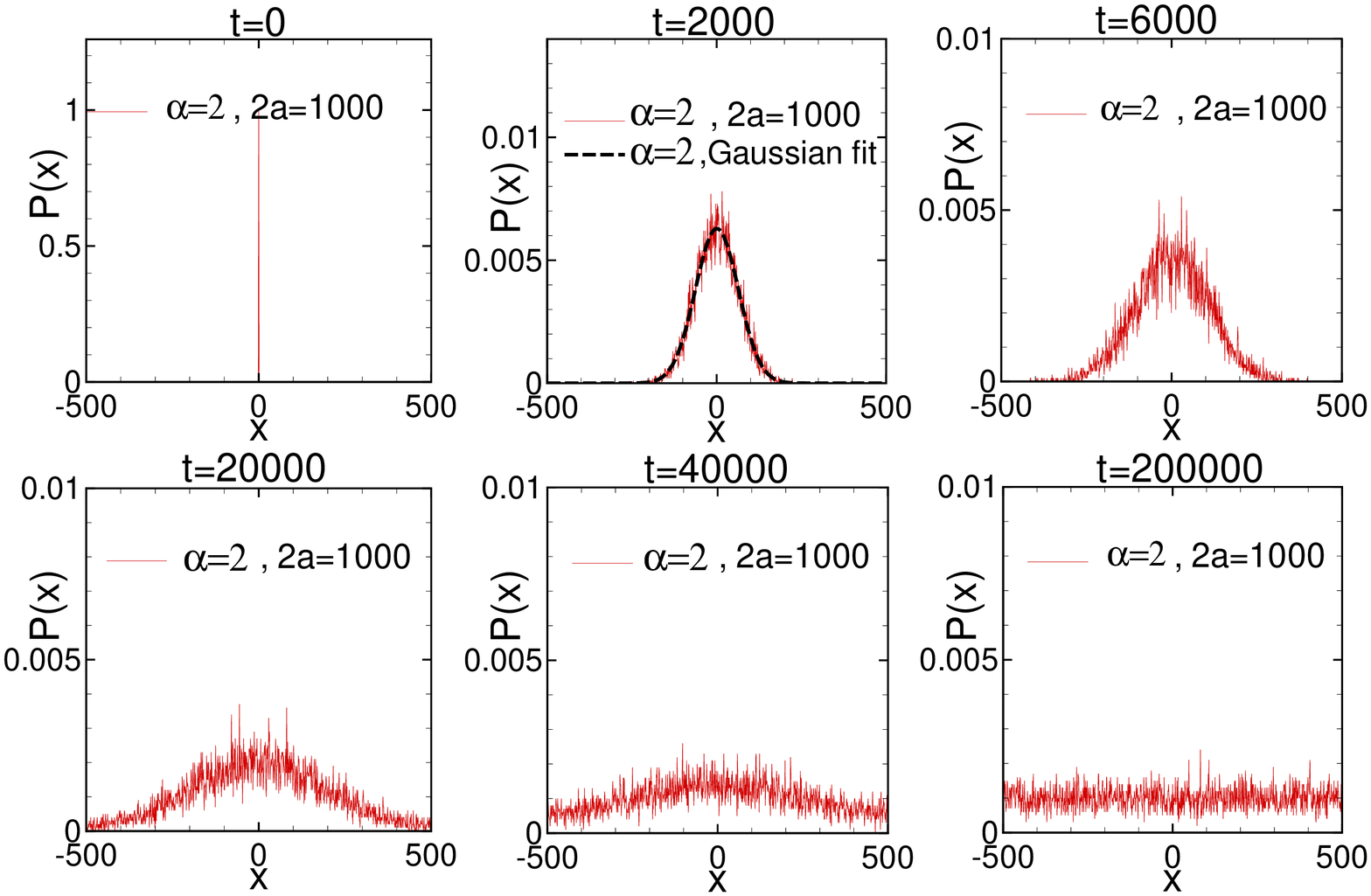}
\caption{Time evolution of the unidirectional probability distribution
function $P(x,t)$ of confined L{\'e}vy flights with $\alpha=0.7$ and
$1.7$, as well as Brownian motion with $\alpha=2$ in a box of edge
length, $2a=1,000$, for $\sigma=\sqrt{2}$. In each case $10,000$
sample trajectories were used to create an individual curve for
$P(x,t)$. Fits (dashed black lines) correspond to stable
distributions as in Eq.~(\ref{char_func}). \label{fig.pdf}}
\end{figure}

We now turn to the probability density function of the process.
Fig.~\ref{fig.pdf}
shows the time evolution of the unidirectional probability density
function $P(x,t)$ for L{\'e}vy flights with stable indices
$\alpha=0.7$ and $1.7$. For comparison we also present graphs for
Brownian motion, $\alpha=2$. Starting with the sharp initial
condition $P_0(x)=\lim_{t\to0}P(x,t)=\delta(x)$, the probability
density function successively broadens, until it reaches the state
of equidistribution, $P_{\mathrm{st}}(x)=1/(2a)$. We observe that
the specific characteristics of the different cases $0<\alpha<1$,
$1<\alpha<2$, and $\alpha =2$ remain discernible up to relatively
long times. Thus up to some 10\% of the overall simulations time, it
is still possible to distinguish the basic shapes of the probability
density functions for these cases. The overall simulations time
needed to reach saturation is, however, dramatically different, as
already observed in the labels shown in Fig.~\ref{fig.pdf}. This Figure
also demonstrates how fitting to a stable law, Eq.~(\ref{char_func}),
yields an approximation of $\alpha$. The fits in this figure show
excellent agreement with simulation data.

\begin{figure}
 \includegraphics{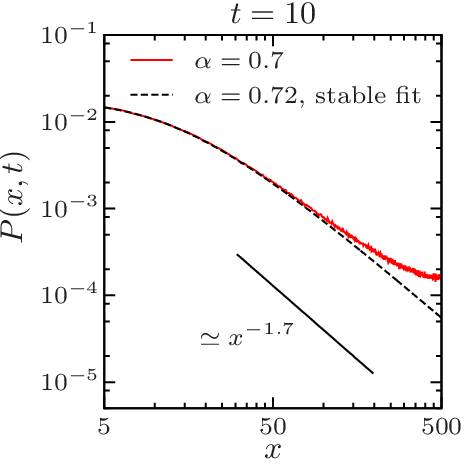}\includegraphics{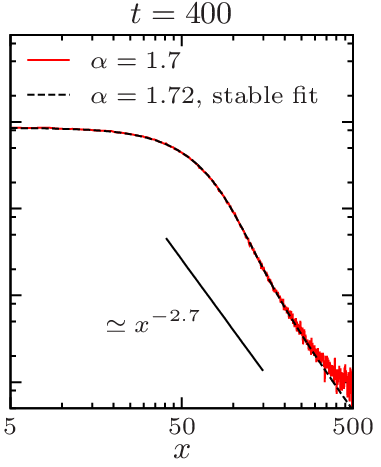}
\caption{Tail analysis for the probability distribution function
$P(x,t)$ of confined L{\'e}vy flights at an early stage of the random walk. The
stable fits (dashed black curve) were performed at the same time instants as in
Fig.~\ref{fig.pdf}, but a larger number of $3\times10^6$ of random walks were
needed to produce precise tail data (continuous red curve).}
\label{fig.pdftail}
\end{figure}

Thus, from fitting probability density functions alone we could
conjecture that the initial dynamics of the searcher starting far
off the edges of the box are essentially unaffected by boundary
conditions: at small times, the spatial distribution can be
approximated by the stable law from an unbounded L{\'e}vy flight.
However, the effect of boundaries is known to be highly non-local if
the random walk is governed by long-tailed jump statistics
(\ref{jld}). Due to the non-negligible probability for extremely
long excursions, the searcher probes the nature of the boundaries
from the very start of its motion. For example, this leads to
modifications on the associated diffusion equations
\cite{Krepysheva,Baeumer}. Fig.~\ref{fig.pdftail} highlights the
specific tail properties of the early-time distribution. Despite the
almost perfect fit quality of stable distributions along virtually
the whole search space, there are considerable deviations in the far
tails near the boundaries. In particular, the distinct heavy-tail
property, Eq.~(\ref{jld}), is already lost at this early stage of
the random walk.

Analyzing the time evolution of moments reveals further distinct
peculiarities of the bounded motion, as we show in the following.

\subsection{Moment analysis}

\begin{figure}
\includegraphics{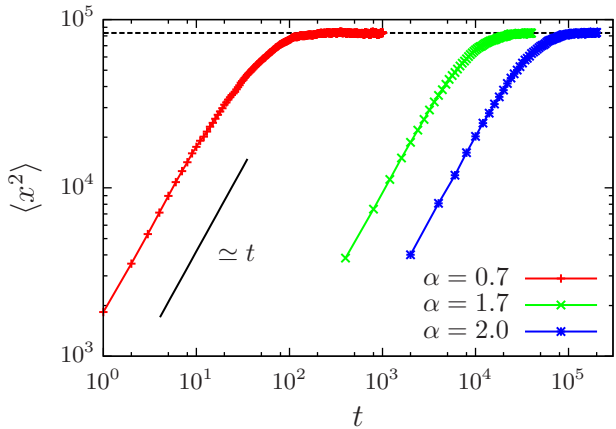}
\caption{One-dimensional mean squared displacement of L{\'e}vy
flights in a box with periodic boundary conditions and size
$1,000\times1,000$ as function of time $t$, for different values of
$\alpha$. Here, 10,000 sample trajectories have been used for the
ensemble average $\langle x^2(t)\rangle$. The horizontal black line
represents the expected stationary value $a^2/3=83,333$, where $a=
500$. Note that the early diffusion process is "normal" in the sense
that $\langle x^2(t)\rangle$ grows linearly in time, despite the
underlying scale-free jump distributions. \label{fig.msd}}
\end{figure}

\begin{figure}
 \includegraphics{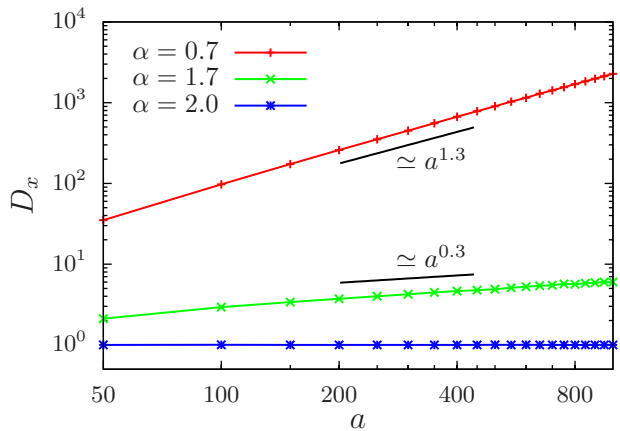}
 \caption{
Diffusion constant versus box size. With the  initially "normal"
growth of the mean squared displacement, one can associate a diffusion
constant $D_x=\langle
x^2(t)\rangle/(2t)$. The latter has a power law dependence on
the edge length $2a$ of the box. Only for bounded Brownian motion, $\alpha=2$,
the (large) system size does not affect the initial diffusion parameters.
Error bars are of the size of the symbols.}
\label{fig.aD}
\end{figure}

First, consider the mean squared displacement $\langle
x^2(t)\rangle$, which is another common approach to assess the area
coverage of the process. While this quantity diverges for a free
L{\'e}vy flight or a L{\'e}vy flight in an harmonic external
potential \cite{report,jespersen}, it is already finite in steeper
than harmonic external potentials \cite{chechkin}. In our finite
area the mean squared displacement is thus a valid measure for the
motion. Figure~\ref{fig.msd} shows the time dependence of the
one-dimensional variance, for different values of the stable index
$\alpha$. The saturation of the mean squared displacement is
comparatively sharp, such that we can read off a typical time scale
$\tau$ from Fig.~\ref{fig.msd}, see below. In the initial diffusion
regime, $t\ll\tau$, all graphs depict a linear time dependence,
independently of $\alpha$. This highly contrasts the properties of
an unbounded L{\'e}vy flight, where $x^2$ scales like
$t^{2/\alpha}$. In fact, from measuring the mean squared
displacement alone one could assume the motion to be confined
Brownian. Thus, the effect of the boundaries goes beyond ensuring
the finiteness of the second moment. Interestingly, the boundary
characteristics also have a distinct quantitative effect on the
initial dynamics, which we see in Fig.~\ref{fig.aD}: the associated
diffusion coefficient $D_x=\langle x^2(t)\rangle/(2t)$ depends on
the size of the system. Enlarging the box increases the (finite)
variance of individual jump lengths, thereby increasing $D_x$.

\begin{figure}
\includegraphics{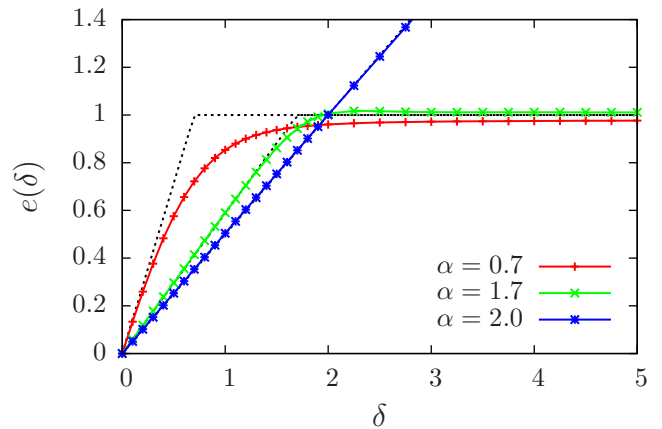}
\caption{Scaling exponent $e(\delta)$ for moments $\langle |x|^\delta
\rangle$ of arbitrary order $\delta>0$. The exponent is obtained by
fitting to a power law, $\langle |x(t)|^\delta \rangle\simeq
t^{e(\delta)}$. The theoretical prediction (black dotted lines,
Eq.~(\ref{moments})) is valid only in the asymptotic limit $1\ll
t\ll\tau\simeq a^{\alpha)}$. For this simulation, $2a=10^5$,
$t\in\{10,...,30\}$. The calculation of higher order moments
exploits the far tails of the distribution, requiring a relatively
large number of $10^8$ trajectories.} \label{fig.moments}
\end{figure}

To get the full picture we compute arbitrary moments of absolute displacements
from our simulations. We generally find an initial power law time
dependence, $\langle |x|^\delta(t)\rangle\simeq t^{e(\delta)}$, before
the occurrence of saturation. More precisely, from the data for the
scaling exponent $e(\delta)$ displayed in Fig.~\ref{fig.moments},
we expect that in the asymptotic limit $1\ll t\ll\tau$
\begin{align}
\langle |x|^\delta(t)\rangle \simeq
\begin{cases}
 t^{\delta/\alpha}, & 0<\alpha<2,\quad 0<\delta<\alpha,\\
 t, &  0<\alpha<2,\quad \delta\geq\alpha,\\
 t^{\delta/2}, &\alpha=2, \quad\delta>0.
\end{cases}
\label{moments}
\end{align}

We note that this type of moment scaling is actually a
familiar one in the
theory of L{\'e}vy flights. It also occurs when truncating the
distribution~(\ref{jld}) for the jump lengths at some fixed
distance~\cite{Mantegna,Koponen,Nakao}, or
when
considering the
effects of a finite ensemble~\cite{Chechkin2}. Roughly speaking, the
truncation distance is
analogous to the largest value drawn from a finite ensemble; in
our system this role is played by the finite edge length of the box. In all
such cases, the initial dynamics
resemble
an ordinary, unbounded L{\'e}vy flight, albeit with finite moments of the
type~(\ref{moments}).

In particular, we conclude that the second moment of bounded
motion mimics ``normal'' diffusion, but the associated diffusion
coefficient \emph{must} depend on the system size parameter $a$: at
early times, moments with $0<\delta<\alpha$ are adopted from the
unbounded L{\'e}vy flight and are thus independent of $a$. At late
times, all moments saturate at $\langle |x|^\delta(\infty)\rangle
\simeq a^\delta$. The typical time scale for the interaction with
the boundaries therefore scales as
\begin{equation}
 \tau \simeq a^\alpha,\quad\text{ but also }\quad \tau \simeq a^2/(2D_x).
\label{scale}
\end{equation}
This necessarily implies that
\begin{equation}
 D_x=D_x(a)\simeq a^{2-\alpha}.
\end{equation}
The above relation is in nice agreement with our power law fits in
Fig.~\ref{fig.aD}. The exponent also consistently coincides with the
one found for the dependence on the truncation parameter in
truncated, but unbounded,  L{\'e}vy flights as anticipated in
Ref.~\cite{Nakao}.

\subsection{Mean first passage time}

\begin{figure}
\includegraphics{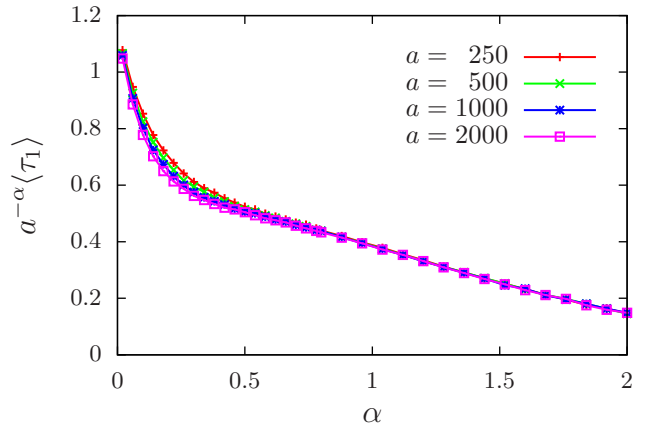}
\caption{Rescaled mean first passage time of confined L{\'e}vy flights as
function of
the stable exponent $\alpha$, for different box sizes. Here, $10^4$-$10^5$
sample trajectories
have been used for each data point.}\label{fig.mfpt}
\end{figure}

We finally analyze the mean first passage time
$\langle\tau_1\rangle$ of our confined L{\'e}vy flights to the
boundary of the area. On a semi-infinite domain L{\'e}vy flights
show the universal Sparre Anderson scaling $\simeq \tau_1^{-3/2}$ of
the distribution of first passage times, so the mean first passage
time $\langle\tau_1\rangle$ diverges \cite{chechkin,koren}. On our
finite domain it converges and depends crucially on $\alpha$. We
determined the mean first passage time for L{\'e}vy flights with
different values of $\alpha$ and for different box sizes ($a=250$,
$500$, $1000$ and $2000$). From scaling arguments we would expect
that the mean first passage time grows like
$\langle\tau_1\rangle\simeq a^{\alpha}$ as function of the initial
distance from the absorbing boundary, see Eq.~(\ref{scale}).
Fig.~\ref{fig.mfpt} depicts the rescaled mean first passage time
$a^{-\alpha}\langle\tau_1\rangle$ as function of $\alpha$. We
observe that the scaling relation holds provided we can approximate
the L{\'e}vy flight by a continuous diffusion process. Conversely,
by choosing small values for $a^\alpha$ we generate a random walker
that escapes the box after only a few number of steps and the
universal scaling~(\ref{scale}) breaks down. From the monotonic
nature of the graphs we conjecture that -- for certain fixed
parameters $\sigma$ and $a$ -- one might find an optimal value for
the tail parameter $\alpha$ which minimizes the mean first passage
time.

We note that the analogous first passage time problem in one
dimension has been treated elsewhere analytically and numerically
\cite{Buldyrev,Zoia}. A generalization to our two-dimensional
problem however is not straightforward: Although we fix the scale
$\sigma=\sqrt2$  for individual jump distances $r$ , displacements
are still affected by random jump directions $\theta$. One can argue
that in effect, the scale of displacements along the $x-$, $y-$ or
radial coordinate is \emph{not} fixed, but has a non-trivial
relation to the tail parameter $\alpha$. Consequently, such
considerations raise the general question of how to appropriately
fix a jump length scale for variable stable exponents in
multi-dimensional L{\'e}vy flights.

\section{Conclusions}

Area coverage of stochastic processes is an important criteria for
the efficiency of search or spreading processes. Here we
investigated the time evolution of the time average of the area
coverage in finite territories of L{\'e}vy flights. As measures we
used the fractal dimension, which relaxes to two at sufficiently
long times, and moments of displacements of the L{\'e}vy flight.
Consistently, we found that both quantities show a relatively
immediate saturation behavior, such that a clear time scale can be
determined for reaching full area coverage. However, this time scale
crucially depends on the value of the stable index $\alpha$: when
$\alpha$ approaches the limiting value two, this time scale
increases significantly. As expected, smaller values for $\alpha$,
allowing for longer jump lengths effect more efficient area coverage
and thus search efficiency. In general, the number of steps
necessary to reach full area coverage is relatively large. Thus
searchers using this type of strategy will likely not be able to
achieve absolute certainty to find a given target randomly. A field
of vision, as introduced in the search literature, will mellow this
problem and render L{\'e}vy flights a very efficient search strategy.

The analysis of the area coverage is completed with the study of the
equilibration behavior of the probability density function of the
process, allowing one to distinguish different domains of the stable
index $\alpha$ (smaller/larger than one) up to relatively long
times. Moreover the mean first passage time $\tau_1$ scales with the
index $\alpha$ and the system size approximately in the scaling fashion
$\langle\tau_1\rangle\simeq a^{\alpha}$.

From the data analysis point of view, we find that studying one
single type of above measures is usually not enough to identify a
given motion as a bounded L{\'e}vy flight. The fractal dimension
does have a characteristic behavior for long-tailed jump lengths,
but quite extensive data is necessary to achieve reliable results
from a box counting algorithm. Although fitting the spatial
distributions by stable laws might agree nicely at early stages of
the process, their characteristic heavy tails are quickly reshaped
by interactions with the boundaries. While the mean squared
displacement indicates normal diffusion, the associated diffusion
constant turns out to depend on the system size and lower order
moments indicate super-diffusion. Finally, an analysis based on
first passage times only yields decisive results if data on a
variety of system sizes is available. We conclude that usually a
combination of methods is necessary to unambiguously determine the
L{\'e}vy flight nature of a given process.

\acknowledgments

Financial support from the Academy of Finland (FiDiPro scheme) and the
CompInt graduate school at TUM are gratefully acknowledged.

\end{document}